\documentclass[prx,twocolumn,american,aps,revmod,superscriptaddress,nofootinbib,showpacs]{revtex4-1}
\usepackage[english]{babel}
\usepackage[T1]{fontenc}
\usepackage{setspace}
\usepackage{outlines}

\usepackage[colorinlistoftodos]{todonotes}
\usepackage[colorlinks=true, allcolors=blue]{hyperref}
\usepackage{amsmath}
\usepackage{xcolor}
\usepackage{placeins}
\usepackage{booktabs}
\usepackage{float}
\usepackage{tabularx} 
\usepackage{graphicx} 
\usepackage{array}
\usepackage{adjustbox} 
\usepackage{amsmath,amsbsy} 
\usepackage{comment}
\usepackage{booktabs}
\usepackage[normalem]{ulem}
\usepackage{multirow}

\makeatletter
\renewcommand{\l@section}{\@dottedtocline{1}{1.em}{1.9em}}
\renewcommand{\l@subsection}{\@dottedtocline{2}{2.0em}{1.5em}}
\renewcommand{\l@subsubsection}{\@dottedtocline{3}{3.em}{1.5em}}
\makeatother

\begin{document}

\newcommand{\new}[1]{#1}
\newcommand{\old}[1]{}

\title{Physics-Informed Transformation Toward Improving the Machine-Learned NLTE Models of ICF Simulations} 
\author{Min Sang Cho}
\email{cho28@llnl.gov}
\author{Paul E. Grabowski}
\email{grabowski5@llnl.gov}
\author{Kowshik Thopalli}
\author{Thathachar S. Jayram}
\affiliation{Lawrence Livermore National Laboratory, 7000 East Avenue, Livermore, CA 94550, USA}

\author{Michael J. Barrow}
\affiliation{Lawrence Livermore National Laboratory, 7000 East Avenue, Livermore, CA 94550, USA}
\affiliation{Stanford University, Stanford, CA 94305, USA}

\author{Jayaraman J. Thiagarajan}
\author{Rushil Anirudh}
\author{Hai P. Le}
\author{Howard A. Scott}
\author{Joshua B. Kallman}
\author{Branson C. Stephens}
\author{Mark E. Foord}
\author{Jim A. Gaffney}
\author{Peer-Timo Bremer}
\affiliation{Lawrence Livermore National Laboratory, 7000 East Avenue, Livermore, CA 94550, USA}

\date{\today}
\begin{abstract}

The integration of machine learning techniques into Inertial Confinement Fusion (ICF) simulations has emerged as a powerful approach for 
enhancing computational efficiency. By replacing the costly Non-Local Thermodynamic Equilibrium (NLTE) model with machine learning models, 
significant reductions in calculation time have been achieved. \new{However, determining how to optimize machine learning-based NLTE models in order to match ICF simulation dynamics 
remains challenging, underscoring the need for physically relevant error metrics and strategies to enhance model 
accuracy with respect to these metrics. Thus, we propose novel physics-informed transformations designed to 
emphasize energy transport, use these transformations to establish new error metrics, and demonstrate that they yield smaller 
errors within reduced principal component spaces compared to conventional transformations.}

\end{abstract}
\maketitle

\section{Introduction}
A recent inertial confinement fusion (ICF) \cite{nuckolls1972laser} experiment \cite{zylstra2022burning}
at the National Ignition Facility (NIF) 
achieved fusion ignition, obtaining a greater fusion energy yield than the input laser energy.
Since then, ignition has been repeated several times \cite{tollefson2024us}, and it is expected that this result 
will soon become routine, allowing regular experiments to understand the physics of a self-heating
high energy density plasma \new{and to improve the energy gain.
In order to optimize this gain with respect to the experimental configuration, many variations in design parameters must be 
made. Because each experiment is very expensive and time consuming to plan, simulations are used to develop optimum designs.
Such simulations have a high computational cost, dominated by the Non-local
Thermodynamic Equilibrium (NLTE) model \cite{scott2022using,frank2022bands}. While machine learning (ML) techniques 
have significantly reduced this computational cost by a factor of ten 
\cite{kluth2020deep,Humbird_2020, LEE2020108973, Humbird_2021,VANDERWAL2022100308, wal2022transfer,vander2022extension,vander2023transfer},
it is unclear what the best data representation and error metrics are for fast and 
accurate ICF simulations.
}

\new{Researchers have employed various data processing techniques to enhance the robustness and scalability of models. 
One approach involves using a modified logarithmic transformation to reduce the dynamic range of NLTE data. Furthermore,
Vander Wal et al.\,\cite{VANDERWAL2022100308} further proposed replacing the logarithmic transformation 
with a cube root transformation to address the thresholding effect that artificially inflates low-opacity values for low-Z elements. 
However, the precise impact of these transformations on the final outcomes of ICF simulations has yet to be thoroughly investigated, 
and the potential of other transformations to further improve these outcomes has not been fully explored.}

\new{A more critical issue is the lack of appropriate error metrics used in training ML models, which can result in the oversight 
of significant effects, physically relevant to ICF simulations. 
Kluth et al.\,\cite{kluth2020deep} and subsequent studies \cite{Humbird_2020, Humbird_2021, VANDERWAL2022100308, vander2022extension, vander2023transfer} have employed 
Rosseland and Planck means, as well as integrated emissivity, as physics-based error metrics. While these metrics are suitable 
for evaluating photon energy transmission in optically thick media, optically thin media, and total photon energy emission, respectively, they are 
not resolved by photon energy groups and are not directly applicable to systems without a well-defined radiation temperature, 
such as NLTE regions in ICF simulations.}

\new{
    To address these issues, we introduce physics-informed transformations (PhITs) and 
    corresponding error metrics. The PhITs are bijective data transformations of absorption and 
    emissivity—two outputs of NLTE models—into variables that describe net energy exchange and photon mean free path, 
    which are fundamental physical quantities. This approach enables the development of generalized models, as the 
    transformed data should adhere to physical laws regardless of the conditions. Additionally, error metrics derived from 
    these transformations allow for a quantitative analysis of how data processing affects radiation properties in ICF simulations, 
    providing guidance for models to minimize these errors and emphasize the most important physical phenomena.
    }
\new{In order to provide context for the PhIT transformation,} 
we describe the radiation-hydrodynamics that motivates the development of \new{the} PhIT in Section \ref{radhyd}. 
\new{Also,} Section \ref{ICFSim} details the process by which we generated our training data. 
In Section \ref{pil_form}, we introduce \new{the} PhIT as a tool to support machine 
learning, highlighting its advantages for radiation hydrodynamics simulations. 
Section \ref{pca_anal} offers a physical interpretation of errors generated through 
Principal Component Analysis (PCA) and illustrates how the \new{PhIT optimizes errors with respect to the dimensional size
of the PCA space}. Finally, in Section \ref{summary}, we provide a summary and 
offer insights for further testing of the transformation.

\section{\label{radhyd} Radiation-Hydrodynamics} 
ICF experiments are usually modeled by multi-group radiation hydrodynamics (RH), which entails evolving both properties of radiation 
fields and materials. \new{The energy moment of} the radiation transport equation, which captures the propagation of radiation 
through the plasma, is given by:
\begin{equation} \label{radeq}
    \frac{\partial U_g}{\partial t}+\nabla\cdot \mathbf{F}_g = e_g-a_g, 
\end{equation}
where \(g\) is the index of the photon energy groups with photon energy ranges equal to $(\Delta E)_g$, 
\(U_g\) is the radiation energy density, \(\mathbf{F}_g\) is 
the radiation flux density, $e_g = 4\pi(\Delta E)_g \epsilon_g/h$ is the emission power density,
$a_g = c \alpha_g U_g$ is the absorption power density, 
\(\alpha_g\) is the absorption coefficient, \(\epsilon_g\) is the emission power density per unit frequency per
steradian, $h$ is Planck's constant, and $c$ is the speed of light.
This equation describes how radiation is absorbed and emitted within the plasma. Often, we use Fick's law 
as a closure:
\begin{equation} \label{Fick}
    \mathbf{F}_g=-\frac{c}{3\alpha_g}\nabla U_g.
\end{equation}   
\new{which leads to the radiation diffusion approximation \cite{colvin2013extreme}.} We emphasize that regardless of whether 
one uses this closure, $\alpha_g$ controls the distance that the photons can travel before being absorbed.

Outside of ICF applications (e.g. astrophysics), further simplifications are usually made.
A radiation temperature \new{($T_r$)} is defined as
\begin{equation}
    \new{ \sigma T_r^4 = \pi \sum_g U_g,}
    \end{equation}
where $\sigma$ is Stefan-Boltzmann's constant. If the radiation is approximately Planckian ($U_g \approx B_g = \int_g B_\nu\,d\nu$), 
this temperature evolves according to
\begin{equation} \label{radtempeq}
    \new{\frac{\partial T_r}{\partial t} -  \nabla \cdot \frac{c}{3\alpha_R \pi^{2}} \nabla T_r = \frac{\pi e}{4 \sigma T_r^3} - \frac{c}{4 \pi^{2}} \alpha_P T_r,}
\end{equation}
where the Planck mean absorption coefficient is
\begin{equation}
\alpha_P = \frac{\sum_g \alpha_g B_g}{\sum_g B_g},
\end{equation}
the Rosseland mean apsorption coefficient is
\begin{equation}
    \frac{1}{\alpha_R}=\frac{\sum_g \frac{1}{\alpha_g}\frac{\partial B_g}{\partial T_r} }{\sum_g \frac{\partial B_g}{\partial T_r}},
\end{equation}
and the integrated emissivity is
\begin{equation}
    e = \sum_g e_g.
\end{equation}
\new{All of these definitions are formulated in terms of summation rather than integration. 
This distinction arises from the nature of the absorption and emissivity data under consideration, 
which are derived from group-averaged datasets. Should the group-averaged data be computed in accordance with 
the specified definitions, they would, in essence, produce values that are consistent with those obtained
through integration-based approaches.} 
The energy gained and lost to the radiation field is released or absorbed primarily by the 
electrons in the plasma. As such, the electrons' temperature is coupled to the radiation field via:
\begin{equation} \label{tempeq}
    \new{\rho C_v \frac{\partial T_e}{\partial t} = - \sum_{g} \left( e_g - a_g \right) + S,}
\end{equation}
where \(C_v\) is the specific heat capacity at constant volume, \(\rho\) is the mass density, \new{and} \(T_e\) is 
the electron temperature, respectively. The term \(S\) 
encompasses other sources and sinks of energy (e.g. ion-electron temperature equilibration, fusion, and thermal conduction).
We note that Eqs. \ref{radeq} and \ref{tempeq} are integrated simultaneously and that the 
absorption and emission depend on both plasma and radiation properties.
Detailed information of the remaining radiation hydrodynamics equations are described in Ref. \cite{colvin2013extreme, larsen2017foundations}.
Solving the transport equations outlined above necessitates the use of transport coefficients \(\alpha_g\) and \(e_g\) 
as inputs, which, in turn, demand considerable computational resources for NLTE calculations. 
These NLTE computations can account for up to \(90\%\) of the total solution time.


\section{\label{ICFSim} Data}

\begin{figure*}
    \centering
    \includegraphics[width=.95\linewidth]{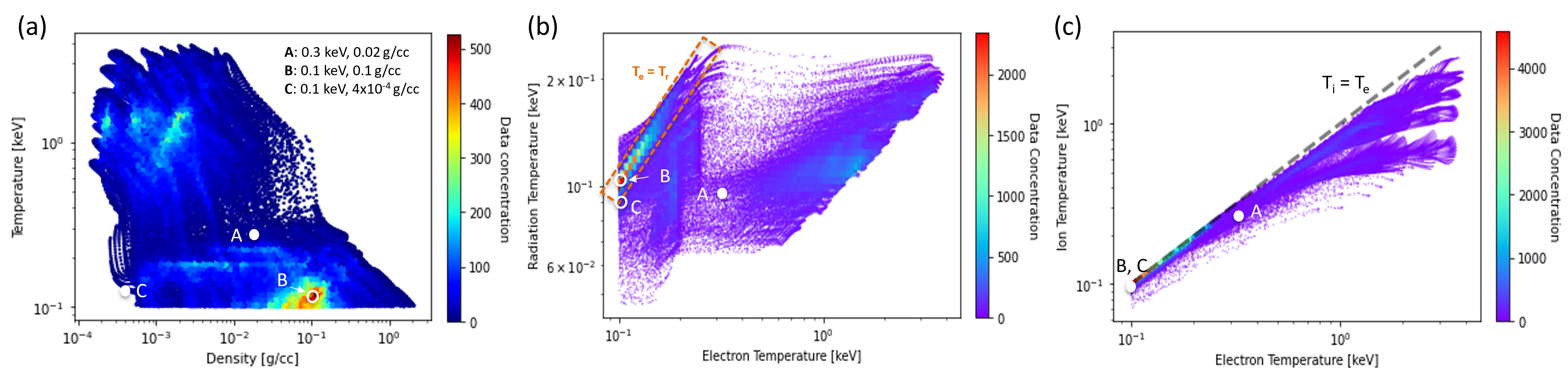}
    \caption{\label{input} (a) Two-dimensional histogram in the temperature-density space, 
    including multiple sets of data with 10 different laser power multipliers, 
    totaling 137,246 data points. Electron temperature vs \new{(b)} radiation temperature and \new{(c)} ion temperature distribution for the data.
    By using training on input data captured from simulations of ICF experiments, we naturally obtain
    input near LTE and reasonable deviations from it.
    }
\end{figure*}

\subsection{Data Structure}
The NLTE model maps electron and ion temperature, density, and radiation fields to 
emissivity and absorption, shown in the Tab. \ref{s-table}.
The plasma mass density ($\rho$), electron temperature ($T_e$), and the ion temperature ($T_i$) are 
scalar quantities while radiation energy density ($\mathbf{U}$) is represented 
as a vector quantity in photon energy groups. The number and sizes of groups
is chosen by the user of the RH code, compromising between computational cost
and accuracy. We used a dataset corresponding to 
60 groups. These unevenly-sized groups span from a photon energy of $0.5$\,eV to $300$\,keV,
with greater resolution where more electron configuration transition energies reside.
The input variable, $\mathbf{U}$ and output variables $\boldsymbol{\alpha}$ and 
$\boldsymbol{\epsilon}$ have this structure.
\begin{table}[t]
    \centering
    \begin{tabularx}{0.5\textwidth}{l>{\centering\arraybackslash}X>{\centering\arraybackslash}X>{\centering\arraybackslash}X}
    \toprule
      & \multicolumn{2}{c}{Inputs} & Outputs \\
    \midrule
    Variables & $\rho$, $T_e$, $T_i$ & $\mathbf{U}$ & $\boldsymbol{\alpha}$, $\boldsymbol{\epsilon}$ \\
    \midrule
    Rank & Scalar & Vector(60) & Vector(60) \\
    \midrule
    Source & RH sims. & RH sims. & NLTE sims. \\
    \bottomrule
    \end{tabularx}
    \caption{\label{s-table} Input and output data utilized for training the Non-Local Thermodynamic Equilbrium (NLTE) surrogate model. 
    The numbers in parentheses adjacent to the vectors denote the dimensionality of the physical quantities. The RH sim. refers to 
    the radiation-hydrodynamics simulation. Our data consisted of 137,246 distinct mappings from these inputs to outputs.}
    \label{table:centered_text}
\end{table}
\subsection{Radiation-hydrodynamics simulations}

Radiation hydrodynamics codes allow us to simulate macroscopic high-temperature systems, in which 
radiation plays an important role in the hydrodynamics.  
These codes integrate the dynamics of fluid motion with the 
transport of radiation to model phenomena such 
as energy deposition, shock wave formation, and the subsequent evolution of the plasma state. They are crucial for 
understanding the dynamics of materials under extreme conditions, such as those found in inertial 
confinement fusion \cite{zylstra2022burning, Abu-Shawareb2022, Abu-Shawareb2024, tollefson2024us}. 
The ability to predict the effects of radiation on plasma dynamics enhances our understanding 
of fundamental physical processes and aids in the design of experiments and the development of advanced technologies in 
fusion energy and related fields.

Input data were created by leveraging a RH code, specifically 
tailored to replicate real ICF scenarios. Utilizing the 2D RH code, KULL \cite{rathkopf2000kull}, 
we captured comprehensive plasma dynamics resulting from laser interactions with a two-dimensional copper target. 
A significant aspect of our methodology involved generating data based on real-world conditions, 
as exemplified by the N101201 laser shot \cite{cho2024}. 
The gradual escalation of the laser intensity over approximately 2 nanoseconds, peaking at 17.5 TW, reflects the 
early phases of ICF (our focus), where the primary mechanism involves the laser heating the hohlraum to initiate X-ray production.
In order to increase 
the size of the dataset, we simulated those shot conditions \new{ten} times, each with a different 
multiplier (0.1, 0.25, 0.5, 0.75, 1, 1.25, 1.5, 2, 5, 10) on the laser power 
directed at a copper hohlraum. Copper allows relatively fast NLTE calculations for 
creating our training dataset, while having 
at room temperature a higher absorbance of light from the
ultraviolet laser than other similar metals (e.g. silver). 

Figure \ref{input}(a) displays the electron temperature and mass density distributions of the 137,246 data points, 
highlighting a range of electron temperatures from 0.1 keV to approximately 3 keV and from $10^{-4}$ to 2.5 g/cc in density. 
Although the distribution is predominantly uniform, there is a noticeable concentration in the regions 
corresponding to a density of 0.1 g/cc and a temperature of 0.1 keV, indicating significant overlap \new{of cases with} the variation in laser \new{multipliers}. Note that our entire dataset had 
$T_e$ > 100 eV, as real-world ICF 
simulations use both emissivity and absorption calculated by the NLTE model only for plasma exceeding temperatures 
of 100 eV and at lower temperatures, pretabulated LTE calculations are thought to be more accurate.

Figures \ref{input}(b) and (c) illustrate the distributions of $T_r$ and $T_i$ in 
relation to $T_e$. Both figures indicate that a substantial portion of the data demonstrates 
thermal equilibrium among the $T_e$, $T_r$, and $T_i$. In a dense 
plasma with a density of approximately 0.1 g/cm$^{3}$ \new{or greater} (e.g. point B in Figure \ref{input}(a)), a simulation time scale of about 2 nanoseconds is sufficient for the establishment of thermal equilibrium. 
However, many data 
points reveal an electron temperature that exceeds both $T_r$ and $T_i$ simultaneously. This 
phenomenon can be attributed to the characteristics of our dataset also, which is derived from simulations of target 
heating by a laser. The data depict a scenario in which electrons are initially heated, resulting in a rise in 
$T_e$ prior to the transfer of energy to the ions through radiation, ultimately leading to a state 
of equilibrium.

\begin{figure*}
    \includegraphics[width=.95\linewidth]{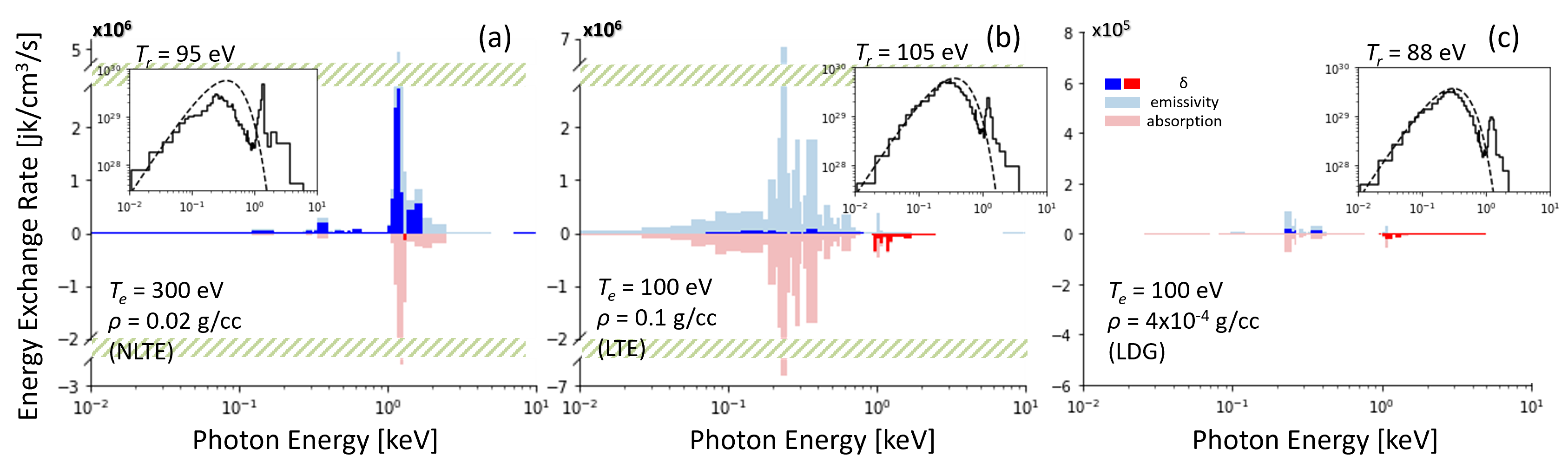}
    \caption{\label{delta_shape} 
    Examples of $\delta_g$, $a_g$, and $e_g$ values are presented for three distinct plasma conditions 
    in the dataset: (a) a NLTE state — Point A in Fig. \ref{input}; (b) a LTE state — Point B; and (c) a low-density region (LDG) 
    — Point C. Positive (dark blue) and negative (dark red) values of \(\delta\) indicate the net radiative energy exchange 
    (emitted minus absorbed radiation energy). Particularly in the photon energy range of \(1 \times 10^{-2}\) to \(1\) 
    keV in condition (b), \(\delta\) approaches zero despite high emissivity and absorption values. The summation of 
    \(\delta\) over photon energy groups yields values of \(8.4 \times 10^6\), \(-1.3 \times 10^6\), and \(1.6\) for 
    conditions (a), (b), and (c), respectively. The magnitudes of $\delta_g$ are larger in NLTE zones and small in 
    LTE zones or at low densities. \new{\textbf{Inset:} The radiation fields for each point with units of 1/cm$^{2}$/s/str 
    in the same photon energy groups. 
    The dashed line in the graph transforms the radiation field energy into a temperature metric, depicting the 
    Planckian distribution for that \( T_r \).} 
    }
\end{figure*}

The radiation fields at three specific points are shown in insets of Figure \ref{delta_shape}. Rather than tracking the
entire spectrum or condensing it to a single scalar (i.e. radiation temperature), our RH simulations tracked the 
radiation energy in 60 photon energy groups.  While the general shape 
adheres to the Planck function, with a non-thermal component near 1 keV due to the L-band emission of Cu, the presence of high-energy photons extending up 
to $10$ keV underscores deviations from standard Planckian distributions at certain temperatures.

\subsection{NLTE Simulations}

The NLTE code is designed to model the population kinetics 
of atomic species in plasma environments where local thermodynamic equilibrium cannot be assumed. This code 
incorporates atomic data to accurately model the distribution of energy levels of ion configurations, 
along with transition rates between them. The first component of the NLTE code is population kinetics, which involves solving the rate equations for the populations 
of atomic energy levels while incorporating both radiative and collisional processes. This approach enables the 
determination of non-equilibrium populations. The second component pertains to spectral synthesis, where the code calculates the opacity and 
emissivity contributions from various processes, including bound-bound, bound-free, and free-free transitions, 
as well as scattering processes. By integrating these components, the NLTE code provides a comprehensive framework 
for analyzing \new{radiative properties of the plasma.}

The plasma conditions, specifically 
electron \new{and ion} temperatures, mass density, and radiation field, \new{extracted}
from KULL simulations, are \new{input} into the Minikin code to compute the absorption and emission coefficients. Minikin, 
a GPU-accelerated version of the Cretin NLTE model used in ICF simulations, performs 
detailed calculations of state populations based on predefined energy levels. While this code can 
output emission and absorption spectra with arbitrary photon energy resolution, we only use those
values averaged over the 60 energy groups corresponding to the KULL simulations. 
The atomic data employed combines the screened hydrogenic model for higher energy levels with the detailed 
configuration accounting (DCA) model for lower levels, providing a nuanced approach to understanding plasma 
interactions under NLTE conditions \cite{SCOTT2001689, kluth2020deep}. For this study, we run the code in 
steady-state and zero-dimensional modes for the single material, copper.

\begin{figure}
    \centering
    \includegraphics[width=.95\linewidth]{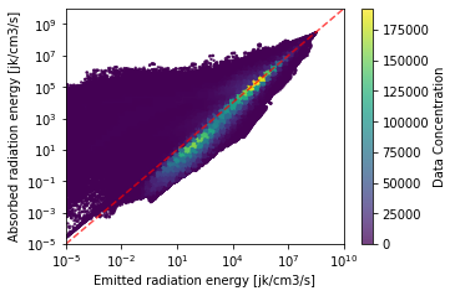}
    \caption{\label{output} Distribution of the radiative power density absorbed and emitted by the plasma. 
    The red line indicates the region where the two values are equal, signifying that the net 
    radiative energy absorption and emission are equal under their correlations such as LTE conditions. 
    The total number of data points in this graph is $8,097,514$, and among these, $868,561$ data 
    points have a 10 $\%$ error margin, accounting for approximately 10.7 $\%$ of the total data points.
    }
\end{figure}

Figure \ref{output} presents the absorption and emissivity values generated by the 
Minikin code, converted into the corresponding radiation power density absorbed by the material and
emitted as radiation, respectively. All 60 photon energy groups are
shown in the same figure. The figure reveals a significant clustering of data points around the equivalence 
line (red dashed line), indicating a strong correlation between emissivity and absorption in the dataset. This 
correlation suggests the potential for effective training by simplifying the data through appropriate transformations.
%
\section{\label{pil_form}Physically-Informed Transformation (PhIT)}

In this section, we introduce a transformation based on two critical physical effects observed in radiation transport: 
(1) the net energy exchange is more important than the emission and absorption individually, 
and (2) mean free paths comparable to the 
system size are more important than subscale and superscale transport. 
The former is exemplified by equations \ref{radeq} 
and \ref{tempeq} where the absorption and emissivity combine to cause energy exchange between radiation and matter.
Thus, we define a new
variable to match this combination: 
\begin{equation} \label{delta}
    \delta_g = e_g - a_g.
\end{equation}
By using $\delta_g$ as an output variable instead of $\alpha_g$ and $\epsilon_g$, the surrogate model
will be natually trained to minimize errors introduced into Eqs. \ref{radeq} and \ref{tempeq}.
Furthermore, the correlation shown in Fig. \ref{output}c between $\alpha_g$ and $\epsilon_g$ will be 
easier for the machine learning algorithm to find since $\delta_g = 0 $ corresponds to perfect correlation.

Figure \ref{delta_shape} provides a visual 
representation of the transformation of absorption and emissivity into the parameter $\boldsymbol{\delta}$
for three example data points. 
In the graph, the light blue bars represent the net emitted radiation power density, $e_g$, 
while the light red bars indicate the net absorbed radiation 
power density, $a_g$. 
The variable $\delta_g$ represents the difference between these two values: a positive value (dark blue) 
indicates the net radiation power density emitted, while a negative value (dark red) indicates the net 
radiation power density absorbed by the plasma. Thus, in Figure \ref{delta_shape}(a), it is evident that 
the plasma at point A is emitting radiation energy, particularly in the 1-2 keV range, 
where significant light emission occurs. Additionally, as the plasma approaches LTE conditions, 
such as in Figure \ref{delta_shape}(b), where the temperature is lower and the density is higher at point B, 
the correlation between emissivity and absorption becomes stronger due to larger matter-radiation coupling, 
leading to their cancellation. Although considerable light emission and absorption occur in the 
0.1-1 keV range, 
the values are nearly equal, resulting in small values of $\delta_g$. Consequently, while the emission and 
absorption rates are higher at point B, from the perspective of total energy exchange, point A is more important. 
In the low-density gas region, both emission and absorption 
rates are minimal, leading to small $\delta$ values, as shown in Figure \ref{delta_shape}(c). Ultimately, 
$\delta_g$ can be regarded as a physical quantity that indicates the importance of energy exchange 
between matter and radiation.

The variable $\alpha_g$ also appears seperate to $\epsilon_g$ in equation \ref{Fick}. It sets the
radiative diffusion length scale. In a physical simulation, some length scales are more imporatant than 
others. If the photon mean free path is much bigger than the hohlraum or much smaller than the hydrodynamics 
zone sizes, the radiation hydrodynamics equations are insensitive to its precise value. As such, we
argue that the thresholding issues worried about in Ref. \cite{VANDERWAL2022100308} are not relevant 
if such a threshold corresponds to mean free paths greater than the hohlraum size. Thus, we can reduce 
the variability of the outputs by introducing a new output variable defined as   
\begin{equation} \label{taug}
    \tau_g = \sigma\left( \frac{\ln(l_g)-\ln(l_{\max} l_{\min})/2}{\ln(l_{\max}/l_{\min})} \right)
\end{equation}
where $\sigma$ is the sigmoid function, $l_g$ is the mean free path, which is the reciprocal of the absorption $\alpha_g$. 
Note that  $l_{max}$ is the hohlraum size (0.3 cm), and $l_{min}$ is the minimum zone size of the RH 
simulation ($3 \times 10^{-8}$\,cm) in this work. 

\begin{figure}
    \includegraphics[width=.95\linewidth]{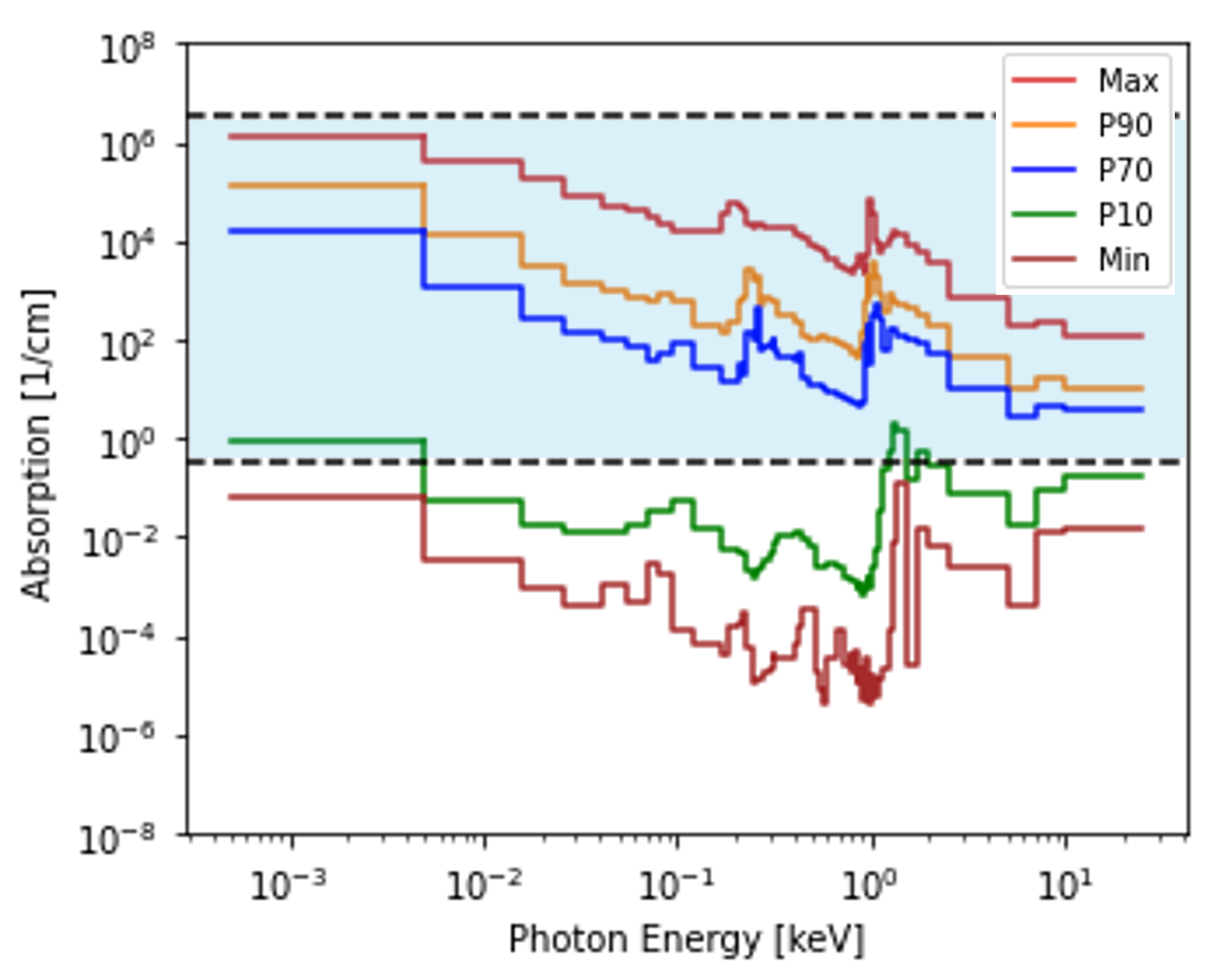}
\caption{\label{tau_shape} \new{Absorption percentiles for datasets.} 
    The shaded blue region indicates regions of this work where absorption values 
    between the maximum threshold defined by the mesh size of 30 nanometers and absorption values that fall below 
    the minimum threshold as set by the hohlraum size of 3 millimeters. `Min' and `Max' correspond 
    to the minimum and maximum absorption values across each photon energy group. The notation `P(number)' specifies 
    the percentile values.
    }
\end{figure}

Figure \ref{tau_shape} displays the absorption data 
across each photon energy group in percentiles, revealing that approximately the bottom 10 percent of data 
points exceed the minimum thresholds set by the conditions of the RH simulation. Thus, it enables us to extract physically 
meaningful regions of absorption by simply transforming
absorption into tau. Additionally, if we had not excluded electron temperatures less than 100 eV, 
a significantly larger portion of the data will fall outside the threshold. 
Consequently, the utility of tau is expected to become even more pronounced.

\section{\label{pca_anal} Transformation Assessment}

For the assessment of the transformations, we employed Principal Component Analysis (PCA), 
a powerful technique for analyzing large datasets with high-dimensional features \cite{strang2019linear}. 
PCA can serve as a surrogate for machine learning, particularly for dimensional reduction models like 
autoencoders, allowing us to analyze the impact of transformations independent of machine learning architectures 
or hyperparameters. To calculate errors, we first perform PCA on the data, 
project each data point into the subspace spanned by a given number of the most important components,
and subsequently calculate the error caused by not including the less important components. 
We initially calculate the error for the original $\epsilon$ 
and $\alpha$ data using this method, and then apply the same approach to data transformed by each respective 
transformation. This process allows us to estimate the size of the latent space needed of each data representation to
obtain a given error.

Real-world datasets often comprise diverse data types, necessitating various 
transformations for effective model training. Given the differing scales associated with physical 
units, Min-Max scaling serves as an essential preliminary step. Moreover, logarithmic and cube root 
transformations are frequently employed to normalize skewed distributions. 

In this section, we aim to demonstrate the effectiveness of the PhIT method in comparison to the 
\new{logarithmic and cube root} transformations, both of which 
are widely used for enhancing training efficacy \cite{box1964analysis, yeo2000new}. 
This comparison will serve to validate the impact of PhIT on model performance. Note, in Ref. \cite{VANDERWAL2022100308}
the logarithmic transform used was $x\to \log(1+x)$, slightly different from here.

\subsection{Error Metrics \label{errmetricssec}}
\renewcommand{\arraystretch}{1.3} 

\begin{table*}[t]
    \centering
    \begin{tabular}{>{\centering\arraybackslash}m{0.1\textwidth} 
                    >{\centering\arraybackslash}m{0.15\textwidth} 
                    >{\centering\arraybackslash}m{0.15\textwidth} 
                    >{\centering\arraybackslash}m{0.15\textwidth} 
                    >{\centering\arraybackslash}m{0.15\textwidth} 
                    >{\centering\arraybackslash}m{0.15\textwidth}}
    \toprule
    & \raisebox{5 pt}{Planck Mean} & \raisebox{5 pt}{Rosseland Mean} & Integrated Emission & \raisebox{5 pt}{$\delta$} & \raisebox{5 pt}{$\tau$} \\
    \midrule

    \multirow{2}{*}{Formula} & \raisebox{-7 pt}{$\frac{|\alpha_P- \alpha_P^\textrm{apx}|}{\alpha_P}$ } & \raisebox{-7 pt}{$ \frac{|\alpha_R - \alpha_R^\textrm{apx}|}{\alpha_R} $}  & \raisebox{-7 pt}{$\frac{|e-e^\textrm{apx}|}{e}$} & \raisebox{-2 pt}{\scriptsize $L_1$: $\sum_g|\delta_g-\delta_g^\textrm{apx}|^{ }$} & \raisebox{-2 pt}{\scriptsize $\sum_g|\tau_g-\tau_g^\textrm{apx}|^{ }$} \\
            &\raisebox{10 pt}{} &\raisebox{10 pt}{} &\raisebox{10 pt}{} &  \raisebox{10 pt} {\scriptsize $L_2$: $\sum_g|\delta_g-\delta_g^\textrm{apx}|^{2}$} & \raisebox{10 pt} {\scriptsize $\sum_g|\tau_g-\tau_g^\textrm{apx}|^{2}$}\\
    
    \small{Related Output Variables} & \raisebox{10 pt}{ \(\boldsymbol{\alpha}\) }
            & \raisebox{10 pt}{\(\boldsymbol{\alpha}\)} 
            & \raisebox{10 pt}{$\boldsymbol{\epsilon}$}
            & \raisebox{10 pt}{$\boldsymbol{\alpha}$, $\boldsymbol{\epsilon}$} 
            & \raisebox{10 pt}{$\boldsymbol{\alpha}$} \\
    
    \small What it \raisebox{13 pt}{emphasizes} & \small Radiation absorption in optically thin media 
            & \small Radiation transmission in optically thick media 
            & \raisebox{15 pt}{\small Total emission}
            & \small Net energy exchange between radiation and matter
            & \small Radiation diffusion within relevant length scales \\

    \bottomrule
    \end{tabular}
    \caption{ \label{errormetrics} Five important physical quantities and the physical phenomena they 
    emphasize. The Planck mean absorption, 
    Rosseland mean absorption, and integrated emissivity metrics were used in previous machine 
    learning-based NLTE models \cite{kluth2020deep,Humbird_2020,Humbird_2021}. In this work, we introduce $L_1$ (difference) and $L_2$ (the square of difference)
    error metrics based on $\delta$ and $\tau$. The superscript `apx' denotes approximate values of a surrogate model.
    }
    \label{table:centered_text}
\end{table*}

When evaluating the error in high-dimensional data using machine learning models, metrics such as $L_1$-
or $L_2$-norm errors are conventionally employed \cite{theocharides2018machine,botchkarev2019new, lu2019error, shi2020multi}. 
However, for absorption and emissivity, which are physical quantities 
describing radiative transport, it is essential to use error metrics that \new{correspond to important physical quantities}
to analyze 
their effects accurately. Notably, error metrics based on Planck mean absorption, Rosseland mean absorption, and integrated emissivity,
which are widely used in radiative transport calculations, are used to evaluate the accuracy of absorption and emissivity \cite{kluth2020deep}.
\new{While the above three metrics are useful for minimizing errors introduced into the solution of Eq. \ref{radtempeq},
we are interested in the solution of Eqs. \ref{radeq} and \ref{tempeq} for ICF systems. 
Thus we propose using the PhIT error metrics based on the variables, $\delta$ and $\tau$.
These can be used to minimize error in the energy exchange between matter and radiation and mean free paths of relevant sizes, respectively.} 
These metrics are summarized in Tab. \ref{errormetrics}

Note that the error metrics based on \(\delta\) and \(\tau\) preserve 
the information of each group by taking absolute values within the summation, whereas other approaches sum the values first 
before calculating the errors, allowing the possible cancellation of errors. While maintaining the absolute value signs outside the summation is conceptually valid for physical 
variables, we propose this definition for machine learning models to facilitate the minimization of errors in group structure. 
Practically, for our dataset, resolving the group-wise errors does not make a qualitative difference in our results.

For clarity and consistency in comparing transformations, each error calculation utilizes 
only the associated outputs, as presented in Table \ref{errormetrics}. For instance, the Planck and 
Rosseland mean errors are solely related to absorption error values. Therefore, the term PhIT in this error 
comparison refers exclusively to the absorption data transformed into the $\tau$ space, while the log and 
cube root transformations also pertain only to the results derived from 
the logarithmic and cube root transformations of the absorption data. The same principle 
applies to the integrated emissivity error, where PhIT denotes the $\delta$ transformed data. 
On the other hand, 
since the $\delta$ transformation incorporates both absorption and emissivity values, the absorption 
values used during the $\delta$ transformation and its inverse transformation are derived from the 
ground truth data to isolate the effects on emissivity. The logarithmic and cube root transformations 
are indicated as errors based solely on the transformed emissivity data. In contrast, for the integrated 
$\delta$ error, both emissivity and absorption are associated outputs, thus the results 
reflect the transformations of both outputs, while the integrated $\tau$ error pertains exclusively 
to the results of absorption.

\subsection{Errors introduced by reduction to principle components}

\begin{figure*}
    \includegraphics[width=.95\linewidth]{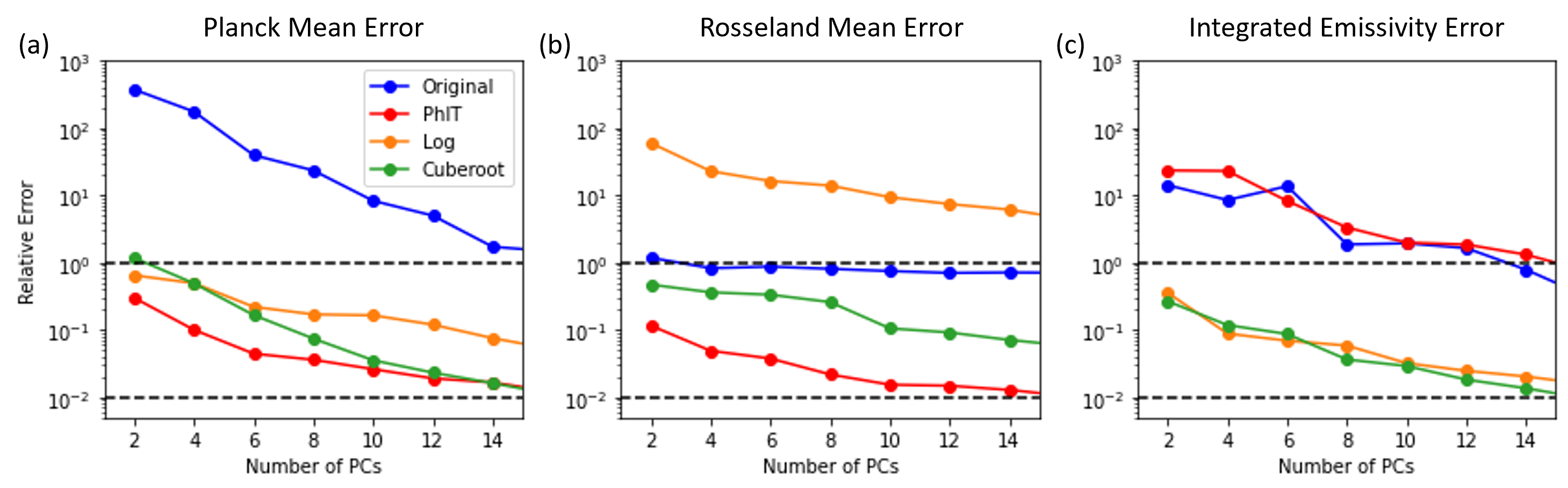}
    \caption{\label{error1} Multiple mean errors introduced by reducing transformed data as a function of number
    of principal components kept in the transformed space. Error metrics shown are (a) Planck, 
    (b) Rosseland mean errors, and (c) integrated emissivity errors.
    Tranforms used are the identity, PhIT, log, and cube root 
    The top and bottom dashed lines correspond to 100\% and 1\% errors, respectively.}
\end{figure*}   

\begin{figure*}
    \includegraphics[width=.95\linewidth]{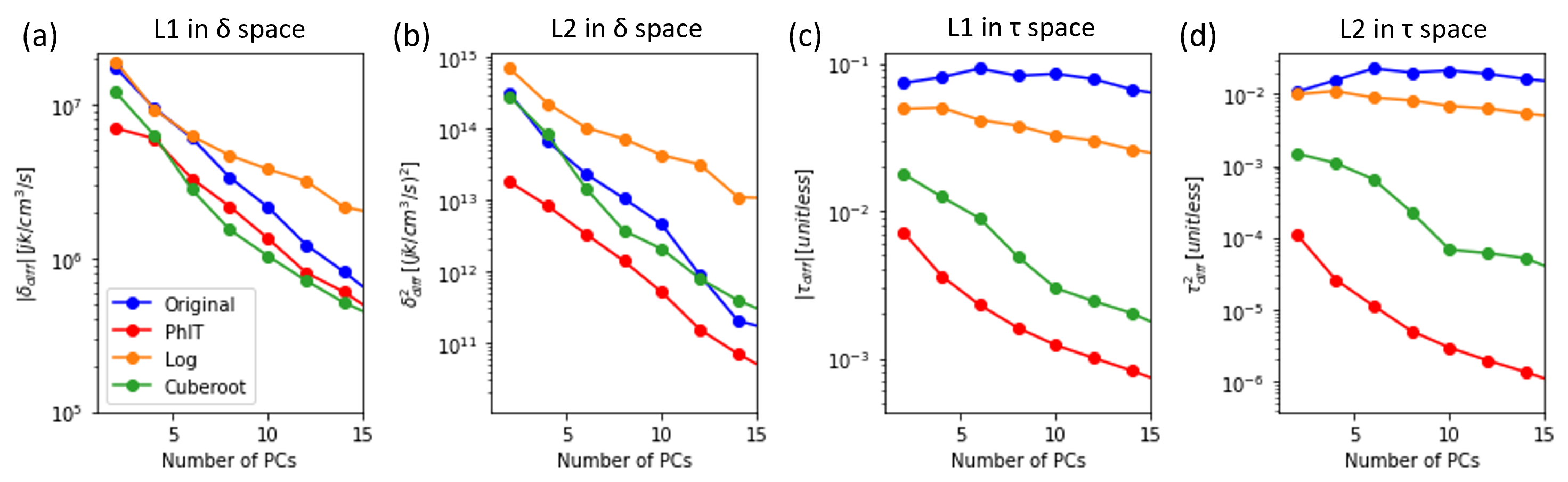}
    \caption{\label{error2} \new{Same as Fig. \ref{error1} except for absolute errors as a function of number
    of principal components kept in the $\delta$ and $\tau$ space. Error metrics shown are (a) the $L_1$, and 
    (b) $L_2$ errors in the $\delta$ space, and (c) the $L_1$, and (d) $L_2$  errors in the $\tau$ space.} 
    }
\end{figure*}

All the error metrics of Sec. \ref{errmetricssec} are shown in Fig. \ref{error1} \new{and Fig. \ref{error2}} as a function of number of 
principal compenents retained averaged over the dataset. The number of 
principal components is analogous to the dimensionality of the latent space, a hyperparameter in machine learning models. 
As the number of principal components increases and approaches the original dimension number, the information loss from the 
dimensionality reduction process decreases, leading to a reduction in error.

Interestingly, Fig. \ref{error1} indicates that transforming the data 
using PhIT significantly improves the Planck and Rosseland mean errors, which are metrics related to absorption.
The PhIT tranformed data reduces mean errors by one to three orders of magnitude 
with respect to the original data. It produces Rosseland mean errors one and three orders of magnitude
smaller than the cube root and logarithmic transformations, respectively. The Planck mean errors are one order of magnitude
smaller than both with small number of principal components and diminishing
difference for the cube root transformation as the number increases.

The principal components most accurately reproduce the data with the largest magnitude and so we expect larger
relative errors where the data are small. In the case of absorption, the data spans 15 orders of magnitude, leading to high variance and large errors for small values. 
Thus, the \(\tau\) transformation, which eliminates both extremely small and large values 
via the sigmoid function, reducing errors. 
While log-transformed and cube root-transformed data also produce similar effects, 
the \(\tau\) transformation demonstrates significantly
greater effectiveness by eliminating emissivity extremes.

On the other hand, the PhIT transformation does not demonstrate significant effectiveness in reducing emissivity errors. 
This observation becomes clear when comparing the $\delta$ transformation with log and cube root transformations. The log and cube root 
transformations serve to narrow the distribution range of data values as previously described, effectively reducing errors in emissivity by 
diminishing relatively small values. In contrast, while the $\delta$ transformation can reduce the distribution range through the 
correlation between emission and absorption, such correlations are not as strong at higher temperatures, which 
also have the largest magnitude emission. The principal components will emphasize these high-temperature, high-emission data points
while not capturing the lower emmision regions, which will have high relative errors. We emphasize that one needs to 
determine for the RH simulation of interest whether one needs low relative or absolute errors. The $\delta$ transformation
adopts an absolute error philosophy because that corresponds to the errors in energy transport.

\begin{figure}[!htbp]
    \includegraphics[width=.95\linewidth]{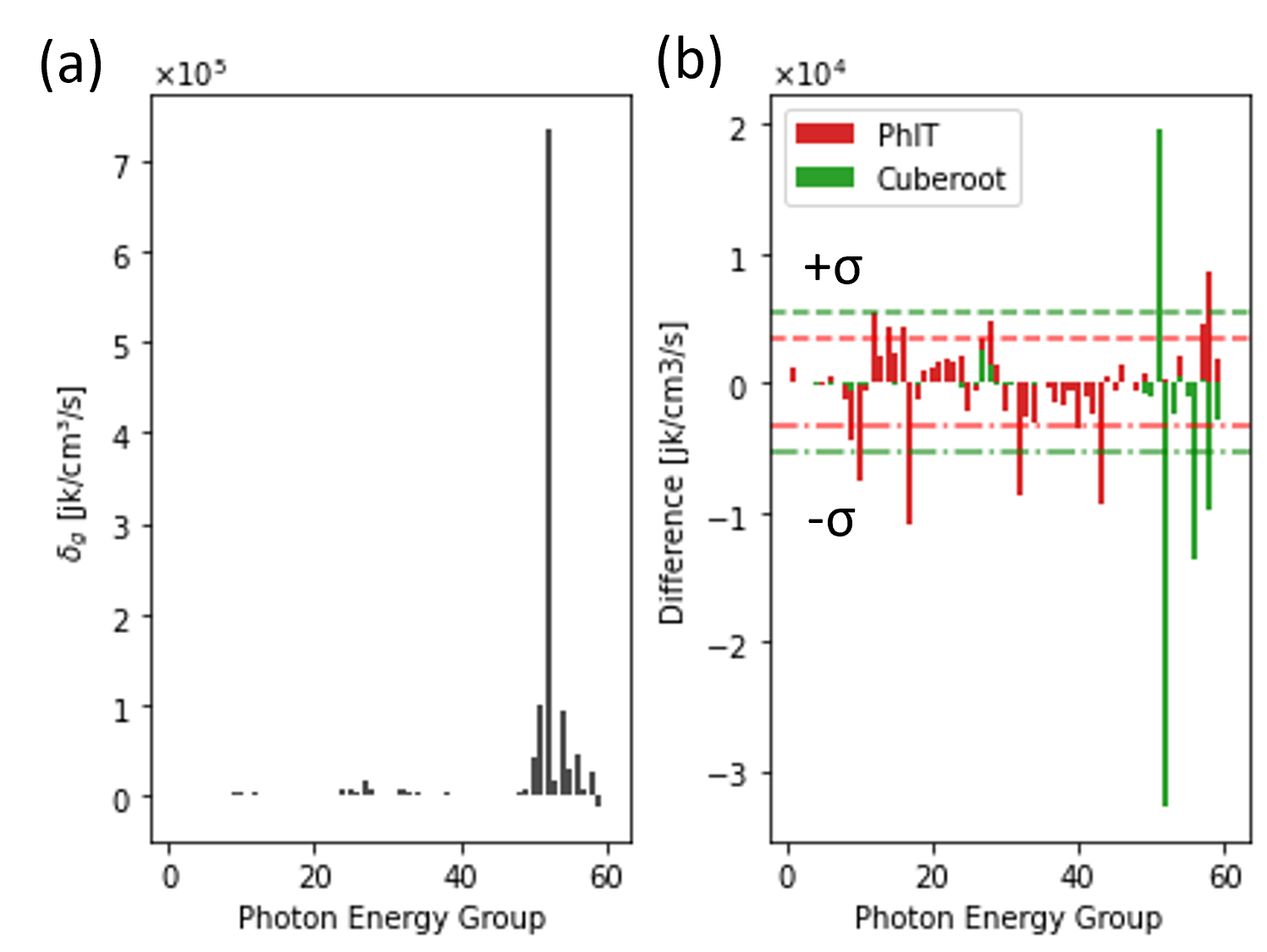}
    \caption{\label{l1_norm} \new{Typical data point with $\delta$-$L_1$ error from the delta transformation being larger
    than the that from the cube root transformation and vice versa with $\delta$-$L_2$ errors.} \new{We show the ground-truth data
    with emission and absorption imbalance concentrated in the higher photon energy groups (a) and
    the errors caused by projecting the data to the 14 largest principal components (b). The $\delta$ transformation
    more evenly distributes errors, leading to its small $L_2$ error despite the larger $L_1$ error. Note, in panel (a), 
    some values of $\delta_g$ are too small to be seen.}}
\end{figure}   
As we see in Fig. \ref{error2}, the PhIT variables generally exhibit the best performance 
in $\delta$ and $\tau$ errors. The exception is that the cube root transformation
leads to slightly lower $L_1$ errors in $\delta$ space. We argued in Sec. \ref{errmetricssec} that 
the $\delta$ and $\tau$ errors are more relevant for multi-group radiation hydrodynamics,
so while it is fortuitous that there are improvements in the Rosseland and Planck mean 
errors, we claim that the ultimate success in applying these transformations will rely on 
the $\delta$ and $\tau$ errors. We show in Fig. \ref{l1_norm} how 
the $L_1$-norm can be smaller for the cube root transformation while the $L_2$-norm is larger: 
the PhIT distributes errors more evenly. Although both the cube root transformation and PhIT 
appear similar 
to the data, the characteristics of the two transformations are differentiated by Fig. \ref{l1_norm}(b). 
Energy exchange frequently occurs within particular photon energy groups in the NLTE region (e.g. Fig. \ref{delta_shape}(a)). Strong radiation emitted 
at specific wavelengths has a decisive impact on the population distribution of the surrounding plasma states, playing 
a crucial role in determining the temperature and density of the plasma in the RH simulation. The use of the $\delta$ transformation 
effectively captures these strong radiation characteristics.

\section{\label{summary} Conclusion}

\new{In this study, we introduced a physics-informed transformation (PhIT) based on specific physical parameters, $\delta$ and $\tau$, 
directly derived from the multi-group radiation hydrodynamic (RH) equations. These parameters are designed to emphasize energy transport, 
enabling more accurate predictions of the resultant plasma dynamics in inertial confinement fusion (ICF) simulations. Additionally, we developed 
PhIT-based error metrics to assess the accuracy of various machine learning models, 
emphasizing the importance of group-resolved energy flow and relevant photon transport length scales in ICF simulations.}

\new{The application of principal component analysis (PCA) facilitated the examination of these transformations. The analysis, 
serving as a surrogate for the latent space of a neural network, allows us to construct errors from each transformed 
parameter and evaluate them with respect to Rosseland and Planck mean absorption, integrated emissivity, 
and the new $\delta$ and $\tau$ metrics. It highlights the ability of $\delta$ and $\tau$ to effectively 
preserve and emphasize the intrinsic complexity of the data, allowing for smaller latent spaces in machine learning-based
non-local thermodynamic equilibrium (ML-NLTE) models compared to other transformations.}

\new{These results underscore the potential of PhIT to enhance the training of machine learning 
models in the field of plasma physics while also providing a robust foundation for future research aimed at optimizing 
data processing techniques. Ultimately, we expect that models trained on PhIT variables will more accurately predict  
plasma dynamics in ICF simulations.}

\section{Acknowledgement}
This manuscript has been authored by Lawrence Livermore National Security, LLC under Contract No. 
DE-AC52-07NA27344 with the US. Department of Energy. The United States Government retains, and the publisher, 
by accepting the article for publication, acknowledges that the United States Government retains a non-exclusive, 
paid-up, irrevocable, world-wide license to publish or reproduce the published form of this manuscript, 
or allow others to do so, for United States Government purposes. We gratefully acknowledge support from 
LLNL's LDRD 22-SI-004.

\FloatBarrier
\bibliography{ref}

\end{document}